\documentclass[article,showpacs,preprintnumbers,amsmath,amssymb]{revtex4}
\usepackage{graphicx}
\usepackage[usenames]{color}

\begin{document}

\title{Unparticle effects in neutrino telescopes}

\author{G. Gonzalez-Sprinberg}
\email{gabrielg@fisica.edu.uy}
 \affiliation{Instituto de F\'{\i}sica, \\ Facultad de Ciencias,Universidad de la
Rep\'ublica \\ Igu\'a 4225, 11400 Montevideo, Uruguay.}

\author{R. Martinez}
\email{rmartinezm@unal.edu.co}
 \affiliation{Departamento de F\'{\i}sica, Universidad Nacional,
Bogot\'a, Colombia}

\author{Oscar A. Sampayo}
\email{sampayo@mdp.edu.ar}

 \affiliation{Departamento de F\'{\i}sica,
Universidad Nacional de Mar del Plata \\
Funes 3350, (7600) Mar del Plata, Argentina}

\begin{abstract}
Recently H.Georgi has introduced the concept of unparticles in order
to describe the low energy physics of a nontrivial scale invariant
sector of an effective theory. We investigate its physical effects
on the neutrino flux to be detected in a kilometer cubic neutrino
telescope such as IceCube. We study the effects, on different
observables, of the survival neutrino flux after through the Earth
and the regeneration originated in the neutral currents. We
calculate the contribution of unparticle physics to the
neutrino-nucleon interaction and, then, to the observables in order
to evaluate detectable effects in IceCUbe. Our results are compared
with the bounds obtained by other non-underground experiments.
Finally, the results are presented  as an exclusion plot in the
relevant parameters of the new physics stuff.
\end{abstract}


\maketitle

\section{Introduction}
The Standard Model (SM) for the  elementary
particles interactions has been successfully tested at the level
of quantum
corrections. In particular high precision and collider experiments
have tested the model and have placed the border line for
physics effects at energies of the order of $1 \rm{TeV}$ \cite{pdg}.
On the other hand, new physics effect in the neutrino sector
have recently received an
important amount of experimental information coming from flavor
oscillation \cite{oscneu}. This fact is the first evidence of
neutrino masses different from zero, and hence, of physics beyond
the SM. In this way, the neutrino sector and in
particular neutrino-nucleon interactions, could be the place where
new physics may become manifest again.
 Although the SM has been
successful to describe the world at short distances, as a low
energy effective theory of phenomena at higher scales, it leaves
several open questions, e.g.: it does not predict the number of
families and the fermions masses,  has  several free parameters, the
mass generation mechanism through the Higgs boson, where its mass is
not predicted, is untested  and still leaves open the hierarchy
problem. In these conditions, it is believed that we should have
some kind of physics beyond the SM, which is called New Physics
(NP)\cite{pdg}. The search of NP proceeds mainly through the
comparison of data with the SM predictions. The experimental way to
look for NP effects in a model independent fashion is to
construct observables that can be affected by this new physics
and then compare the  measurements with the mentioned SM
expectation. Certain types of NP can already be present at the TeV
scale and could participate in neutrino-nucleon interactions. Hence,
these NP effects could possibly become apparent in neutrino
telescopes.

In the other hand it is well-known that scale invariance has been a
powerful tool in several branches of physics and the
possibility of a scale invariant weak interacting sector with the
low energy particle spectrum is not ruled out. In particular, H.
Georgi \cite{georgi} has proposed that a scale invariant sector
which does not imply conformal invariance \cite{grinstein}
with a non-trivial $IR$ fixed point and  coupled to the $SM$
fields through the exchange of particles with a high mass scale
$M_U$ may appear much above the TeV energy scale. Below this energy
scale, this sector induces unparticle operators ${\cal O_U}$ with
a non-integral scale dimension $d_u$ that in turn have a mass
spectrum which looks like a $d_u$ number of massless particles. The
couplings of these unparticles to the SM fields and in particular to
 standard neutrinos (i.e. massless and  left-handed particles)
and quarks are described by the effective Lagrangian \cite{georgi}

\begin{eqnarray}\label{lageff}
{\cal L}_{eff}=\frac{\Lambda^{k+1-d_{\cal U}}}{M^k_{\cal
U}}\overline{f}\gamma_{\mu}(C_{V}+C_{A}\gamma_5)f\;{\cal
O}^{\mu}_{\cal{U}},
\end{eqnarray}

where $\Lambda$ is the energy scale at which scale invariance
emerges, the dimensionless coefficients $C_{V(A)}$ are of order $1$
for neutrinos and quarks and $f$ is the generic fermion spinor. The
operators with lowest possible dimension have the most
important effect in the low energy effective theory regime.
 In Eq.(1) we have only included the vector unparticles operators ${\cal O}^{\mu}_{\cal{U}}$
that couple with left neutrinos and both left and
right quarks.
Note that the left couplings to the neutrinos and quarks are taken equal.

\begin{figure}[t!]
\centering
\includegraphics[width=4in]{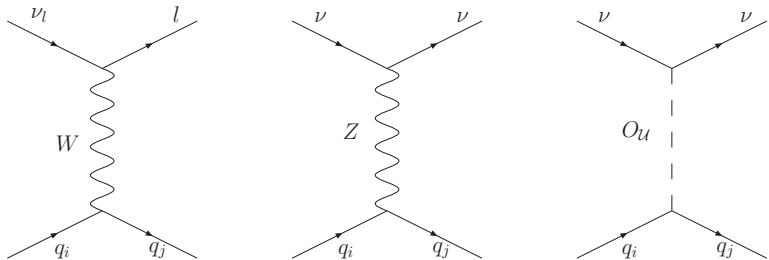}
\caption{\label{fig:diag}  Diagrams contributing to the
 neutrino-nucleon
cross section.}
\end{figure}

In the present work we follow the Georgi's approach where the
Feynman propagators of the unparticle operator $O_u^{\mu}$ is
determined by the scale invariance \cite{georgi},
\begin{eqnarray}\label{propagador}
\int\,d^4x\,
e^{ipx}\,<0|T\Big(O^{\mu}_U(x)\,O^{\nu}_U(0)\Big)|0>=i\,\frac{A_{d_u}}
{2\,sin\,(d_u\pi)}\,\frac{-g^{\mu\nu}+p^{\mu}p^{\nu}/p^2}{(-p^2-i\epsilon)^{2-d_u}}
\, , \label{propagator}
\end{eqnarray}
with
\begin{eqnarray}
A_{d_u}=\frac{16\,\pi^{5/2}}{(2\,\pi)^{2\,d_u}}\,\frac{\Gamma(d_u+\frac{1}{2})}
{\Gamma(d_u-1)\,\Gamma(2\,d_u)} \, . \label{Adu}
\end{eqnarray}
The scale dimension $d_u$ is restricted in the range $1< d_u <2$.
Here,  the condition $d_u>1$ is due to the non-integrable
singularities in the decay rate \cite{davoudias} while
$d_u<2$ is due to the convergence of the integrals \cite{deshpande}.

\section{The Cross Section neutrino-nucleon and the unparticles contribution.}

In this section we consider the effective operator given in
Eq.(\ref{lageff}) and calculate their contribution to the
neutrino-nucleon inclusive cross section:
\begin{equation}
\nu N\rightarrow \mu + \rm{anything},
\end{equation}
where $N\equiv \dfrac{n+p}{2}$ is an isoscalar nucleon. The
corresponding process is pictured in Fig.~\ref{fig:diag} which has
the  SM charged and neutral current diagrams and the unparticle
contribution. For charged currents the calculation is standard and
we use it to compare our results with \cite{gandhi}. For neutral
currents we have included the contributions of unparticles. The
corresponding coupling and propagator are given in equations
(\ref{lageff}) and (\ref{propagador}) respectively.

The SM results for the scattering amplitude for charged
currents muon-neutrino scattering is
\begin{equation}\label{amplcc}
{\cal M}^{CC}_{SM}=-\frac{i g^2}{2(Q^2+M_W^2)} \bar
l\gamma_{\mu} P_L \nu \medspace \sum_{i=D,\bar U} \bar q_j
\gamma^{\mu} P_L q_i,
\end{equation}
and the corresponding differential cross section reads
\begin{equation}\label{disfsigcc}
\begin{split}
 \frac{d\sigma^{CC}}{dxdy}=\frac{G_F^2 s}{\pi}\left(
\frac{M_W^2}{(Q^2+M_W^2)} \right)^2x[Q^{CC}+(1-y)^2 \bar Q^{CC}],
\end{split}
\end{equation}
where for an isoscalar target we have the quark distribution
functions
\begin{equation}\label{districc}
\begin{split}
Q^{CC}(x,Q^2)& =\dfrac{u_v(x,Q^2)+d_v(x,Q^2)}{2}+\dfrac{u_s(x,Q^2)+d_s(x,Q^2)}{2} \\
  & + s_s(x,Q^2)+b_s(x,Q^2),
\\\\
\bar Q^{CC}(x,Q^2)&
=\dfrac{u_s(x,Q^2)+d_s(x,Q^2)}{2}+c_s(x,Q^2)+t_s(x,Q^2).
\end{split}
\end{equation}

Similarly, the SM neutral current amplitude is
\begin{equation}\label{amplncsm}
{\cal M}^{NC}_{SM}=-\frac{i g^2}{2 c_W^2 (Q^2+M_Z^2)} \bar \nu
\gamma_{\mu} P_L \nu \medspace \sum_{i=U,D} \bar q_i
\gamma^{\mu}(g^i_L P_L+g^i_R P_R )q_i,
\end{equation}
where $c_W=\cos \theta_W$, $x_W=\sin^2 \theta_W$,
$g^U_L=1/2-2x_W/3$, $g^D_L=-1/2+ x_W/3$, $g^U_R=-2 x_W/3$ and
$g^D_R=x_W/3$.

From the effective interaction Eq.(\ref{lageff}) between unparticle
and the SM fields we obtain the following four fermion
amplitude
\begin{equation}\label{amplncunp}
{\cal M}_{unp}=\frac{i}{Q^2}\frac{A_{du}}{2 \sin(d \pi)}
\left(\frac{Q^2}{M_Z^2}\right)^{d_u-1}  \bar \nu \gamma_{\mu}
P_L \nu \medspace \sum_{i=U,D}\bar q_i \gamma^{\mu}(c^2_{L}
P_L+c_{L}c_{R}P_R )q_i,
\end{equation}
where the  left ($c_{L}$) and right ($c_{R}$) coupling constants are
expressed in terms of the vector and axial vector coupling
constants:
\begin{equation}
\begin{split}
c_{R}=c_{V}+c_{A} \\
c_{L}=c_{V}-c_{A},
\end{split}
\end{equation}
where
\begin{equation}
c_{V(A)}=\frac{C_{V(A)} \Lambda_u^{k+1-d_u}}{M_u^k M_Z^{1-d_u}}.
\end{equation}

The total neutral current contribution, including the
unparticle contribution, can be written as
\begin{equation}\label{amplnc}
{\cal M}^{NC}=-\frac{i g^2}{2 c_W^2 (Q^2+M_Z^2)} \bar \nu
\gamma_{\mu} P_L \nu \medspace \sum_{i=U,D} \bar q_i
\gamma^{\mu}(\tilde{g}^{i}_L P_L+\tilde{g}^{i}_R P_R )q_i,
\end{equation}
where
\begin{equation}\label{cttnc}
\begin{split}
\tilde{g}^{i}_L=g^i_L-(\delta g) c^2_{L}  \\
\tilde{g}^{i}_R=g^i_R-(\delta g) c_{L} c_{R},
\end{split}
\end{equation}
and
\begin{equation}
\delta g=\frac{A_{d_u}}{\sin(d_u \pi)} \frac{c_w^2}{g^2}
\left(1+\frac{Q^2}{M^2_Z}\right)\left(\frac{Q^2}{M^2_Z}\right)^{(d_u-2)}.
\end{equation}
 The neutral current differential cross section is then
\begin{equation}\label{disfsignc}
\begin{split}
\frac{d\sigma^{NC}}{dxdy}=\frac{G_F^2 s}{\pi}\left(
\frac{M_Z^2}{Q^2+M_Z^2} \right)^2
\sum_{i=U,D}x[\tilde{g}^{i2}_L (Q^{i}+(1-y)^2 \bar Q^{i})  \\
+\tilde{g}^{i2}_R( \bar Q^{i}+(1-y)^2 Q^{i})],
\end{split}
\end{equation}
where the corresponding parton distributions for a isoscalar target
read
\begin{equation}\label{partondistnc}
\begin{split}
Q^U(x,Q^2)=\dfrac{u_v(x,Q^2)+d_v(x,Q^2)}{2}+\dfrac{u_s(x,Q^2)
+d_s(x,Q^2)}{2}  \\+c_s(x,Q^2)+t_s(x,Q^2)
 \\
Q^D(x,Q^2)=\dfrac{u_v(x,Q^2)+d_v(x,Q^2)}{2}+\dfrac{u_s(x,Q^2)
 +d_s(x,Q^2)}{2} \nonumber \\+s_s(x,Q^2)+b_s(x,Q^2)
  \\ \bar
Q^U(x,Q^2)=\dfrac{u_s(x,Q^2)+d_s(x,Q^2)}{2}+c_s(x,Q^2)+t_s(x,Q^2)
 \\ \bar
Q^D(x,Q^2)=\dfrac{u_s(x,Q^2)+d_s(x,Q^2)}{2}+s_s(x,Q^2)+b_s(x,Q^2).
\end{split}
\end{equation}

\begin{figure}
\centering
\includegraphics[angle=270,width=3in,bb= 180 180 550 680]{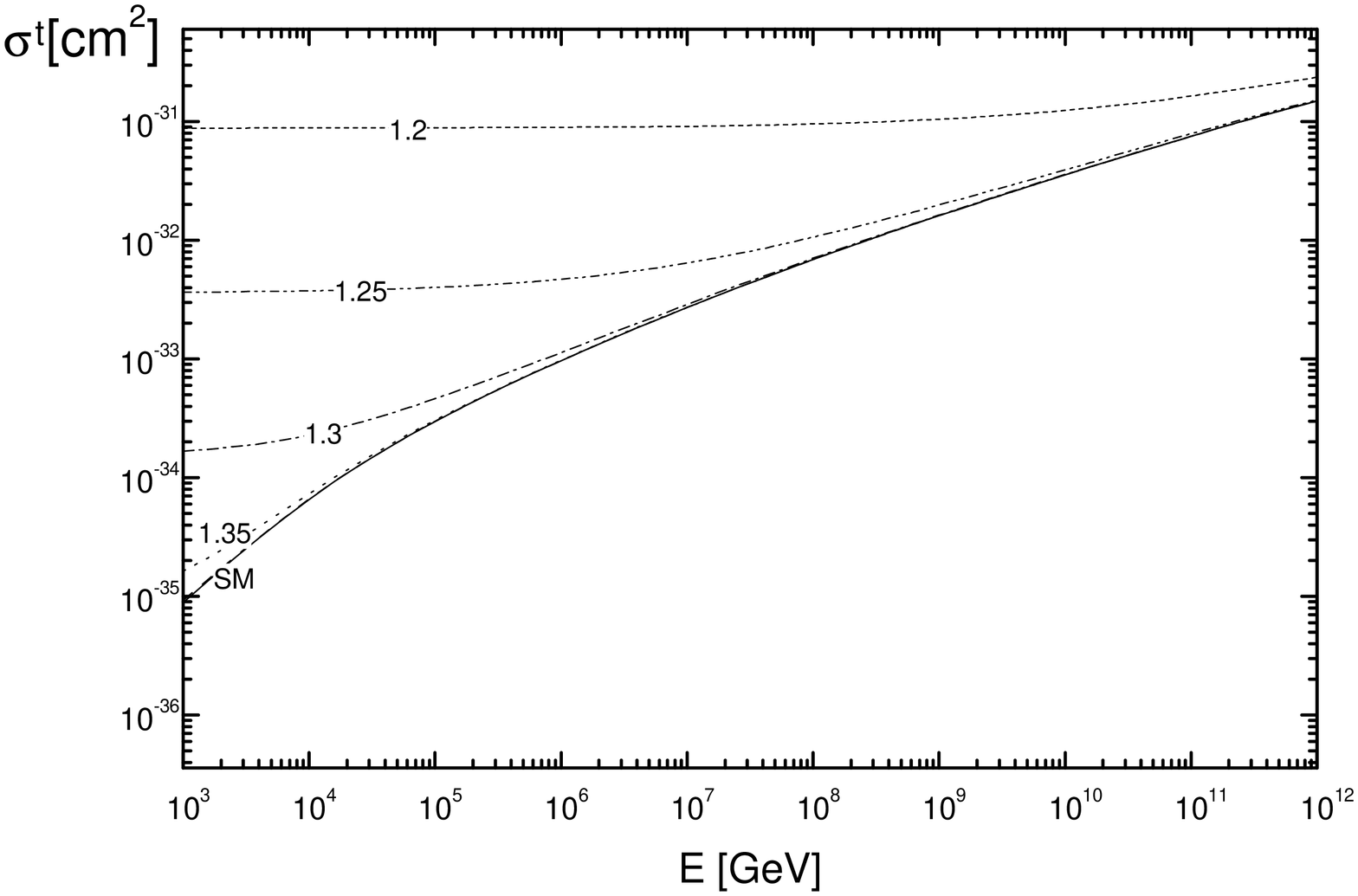}
\caption{\label{fig:sigtot}  Total cross section for the SM
 and for different values of the unparticles dimension $d_u$
and $c_L=c_R=0.01$.}
\end{figure}

In Fig. \ref{fig:sigtot} we show the behavior of the total
cross section ($\sigma^t(E)=\sigma^{CC}(E)+\sigma^{NC}(E)$) with the
neutrino energy for different values of $d_u$ and
$c_{L}=c_{R}=0.01$. We can appreciate a considerable
disagreement with the SM predictions, due to the
unparticle propagator, particularly for low values of $d_u$ and low
neutrino energy. This very disparate behavior do not directly
translate to the neutrino flux due to strong regeneration effects,
as we will see in the next section.
\begin{figure*}
\includegraphics[angle=270,totalheight=8cm,bb= 70 180 550 680]{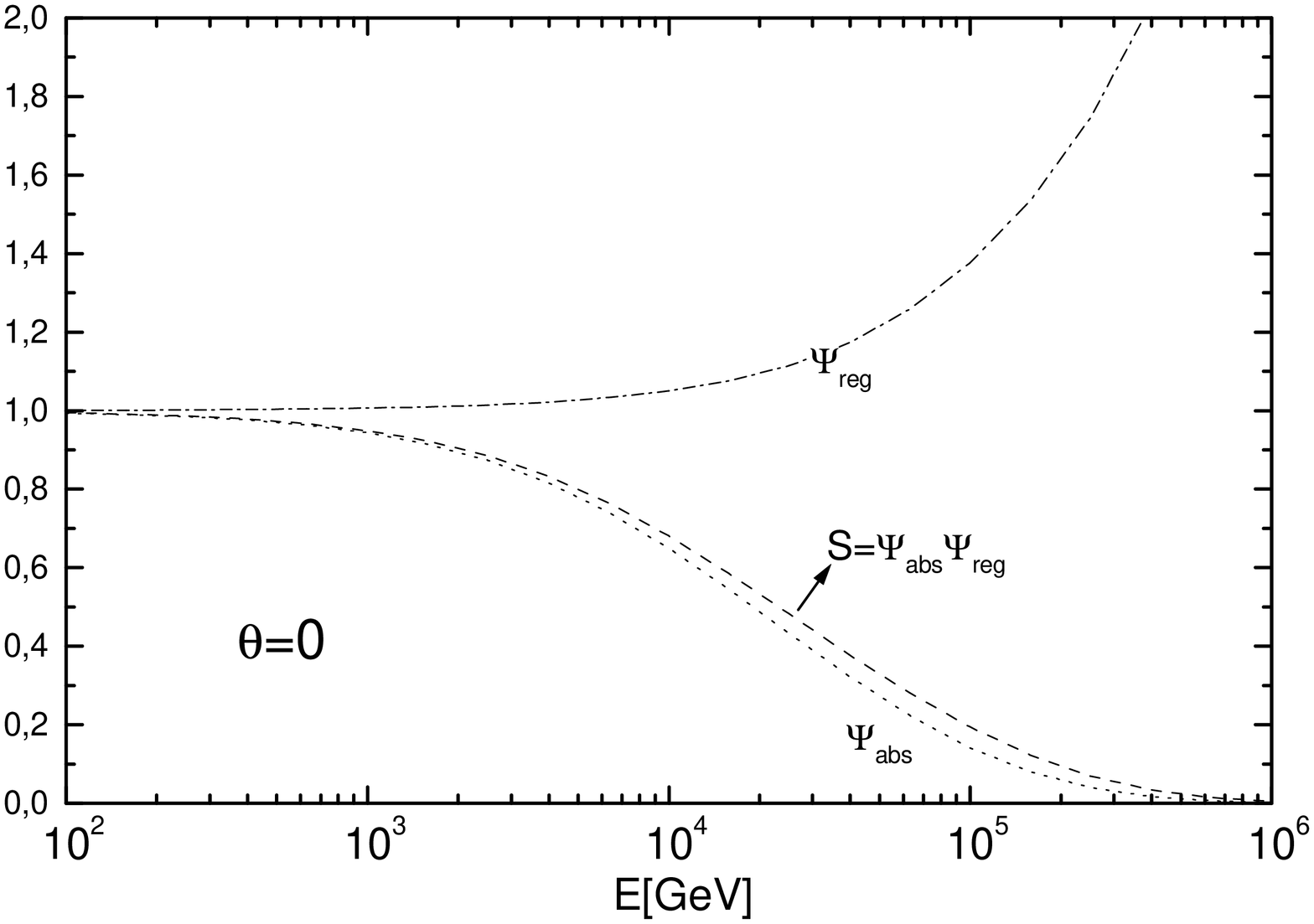}

\includegraphics[angle=270,totalheight=8cm,bb= 70 180 550 680]{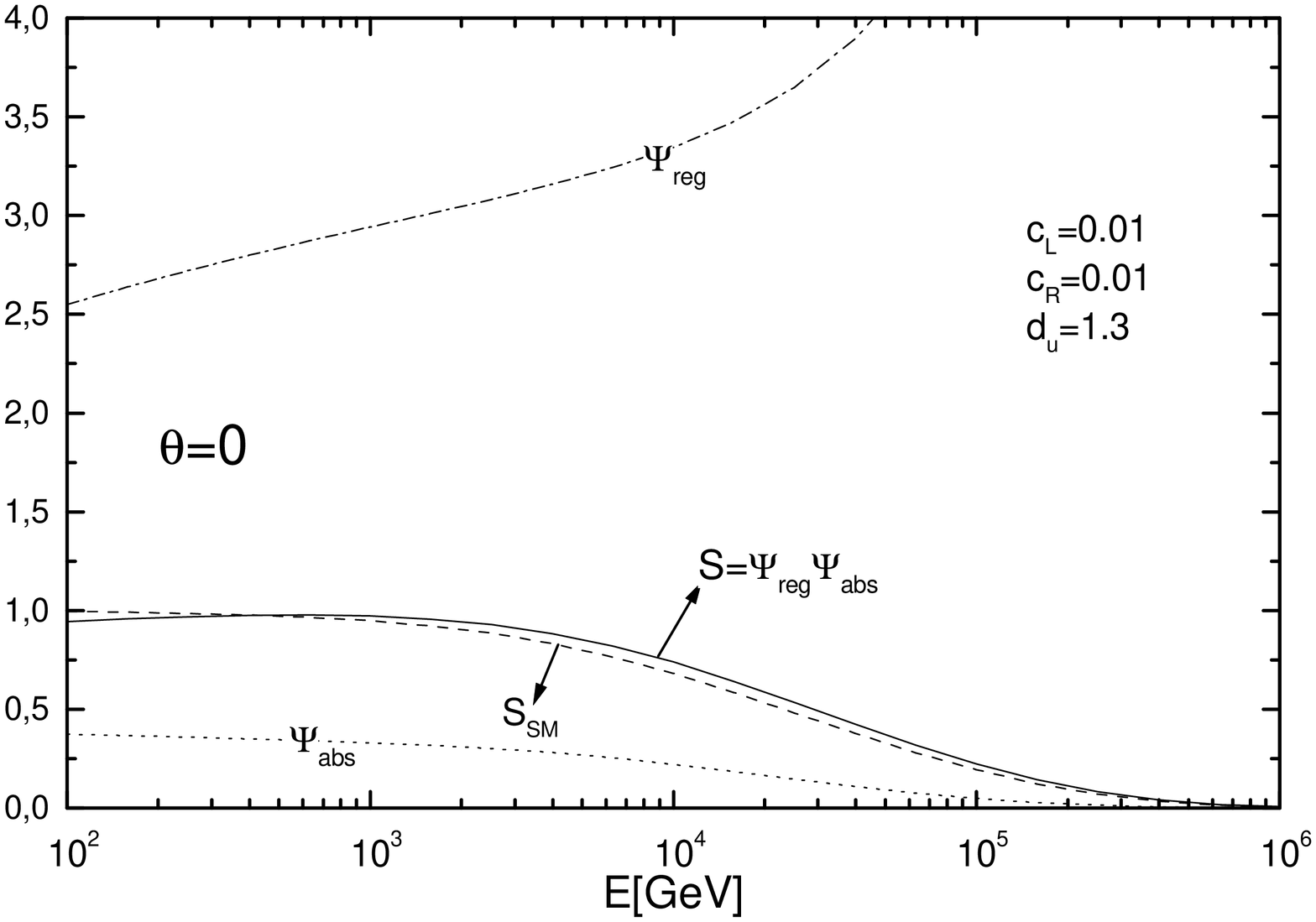}
\caption{\label{fig:psi} Upper: $S$ in the SM. We also include the
factors $\Psi_{abs}$ and $\Psi_{reg}$. Lower: $S$ including the
unparticle contribution for $d_u=1.3$ and $c_L=c_R=0.01$.}
\end{figure*}

\section{The surviving neutrino flux}

The surviving flux of neutrinos of energy $E$, with inclination
$\theta$ with respect to nadir direction, $\Phi(E,\theta)$, is the
solution of the complete transport equation \cite{nicolaidis}:
\begin{equation}\label{ecuaciontransporte1}
\frac{d \ln\Phi(E,\tau')}{d\tau'}=-\sigma^t(E)+\int_E^{\infty} dE'
\frac{\Phi(E',\tau')}{\Phi(E,\tau')} \; \frac{d\sigma^{NC}}{dE},
\end{equation}
where the first term correspond to absorption effects and the second
one to the regeneration. Here,  $\tau=\tau(\theta)$ is the number of
nucleons per unit area in the neutrino path through the Earth,

\begin{equation}\label{tau}
\tau(\theta)=N_A \int_0^{2 R_{\rm E}\cos\theta} \rho(z) dz,
\end{equation}

In order to find a solution for this equation we make the following
approximation \cite{ralston}: we replace the fluxes ratio inside the
integral of the second member by the ratio of fluxes that solve the
homogeneous equation (i.e., only considering  absorption effects)
\begin{equation}\label{ecuaciontransporte2}
 \frac{\Phi(E',\tau')}{\Phi(E,\tau')} \;\; \rightarrow \;\;
 \frac{\Phi_0(E')}{\Phi_0(E)} e^{-\Delta(E',E) \tau'}
\end{equation}
where
\begin{equation}
\Delta(E',E)=[\;\sigma^t(E')-\sigma^t(E)\;] \; \tau'
\end{equation}
Thus, integrating the transport equation we have
\begin{equation}\label{transporte}
\Phi(E,\theta)=\Phi_0(E) e^{-\sigma_{eff}(E,\tau(\theta))
\tau(\theta)},
\end{equation}
where
\begin{equation}
\sigma_{eff}(E,\tau)=\sigma^t(E) - \sigma^{reg}(E,\theta),
\end{equation}
with
\begin{equation}
\sigma^{reg}(E,\theta)=\int_E^{\infty} dE'  \;
\frac{d\sigma^{NC}}{dE} \left( \frac{\Phi_0(E')}{\Phi_0(E)} \right)
\left( \frac{1-e^{-\Delta(E',E) \tau} }{ \tau \Delta(E',E)} \right).
\end{equation}

$\Phi_0(E)$ is the initial neutrino flux considered as isotropic,
$N_A$ is the Avogradro number, $R_{\rm E}$ is the radius of the
Earth, $\theta$ is the nadir angle taken from the down-going normal
to the neutrino telescope and the earth density $\rho(r)$ is given
by the preliminary reference earth model \cite{premm}. It is
important to mention that the solution of the transport equation,
Eq.~(\ref{transporte}) is the first, but quite accurate,
approximation of the iterative method showed in Ref.\cite{perrone}.

In order to illustrate the general behavior of the solution we show
in Fig. \ref{fig:psi}, for $\theta=0$, the factor $S$
\begin{equation}
S=\frac{\Phi(E,\theta)}{\Phi_0(E)}=\Psi_{abs}(E,\theta)
\Psi_{reg}(E,\theta),
\end{equation}
with
\begin{equation}
\begin{split}
\Psi_{abs}(E,\theta)&=e^{-\sigma_{t}(E) \tau(\theta)}  \\
\Psi_{reg}(E,\theta)&=e^{\sigma_{reg}(E,\theta) \tau(\theta)}
\end{split}
\end{equation}
In this figure we show $S$
for the SM and for the unparticles contribution with $d_u=1.3 $ and
$c_L=c_R=0.01$ and we have explicitly included both factors
($\Psi_{abs}$ and $\Psi_{reg}$) to see the compensation between
absorption and regeneration effects.

\begin{figure*}

\includegraphics[angle=270,width=8cm,bb= 150 180 650 680]{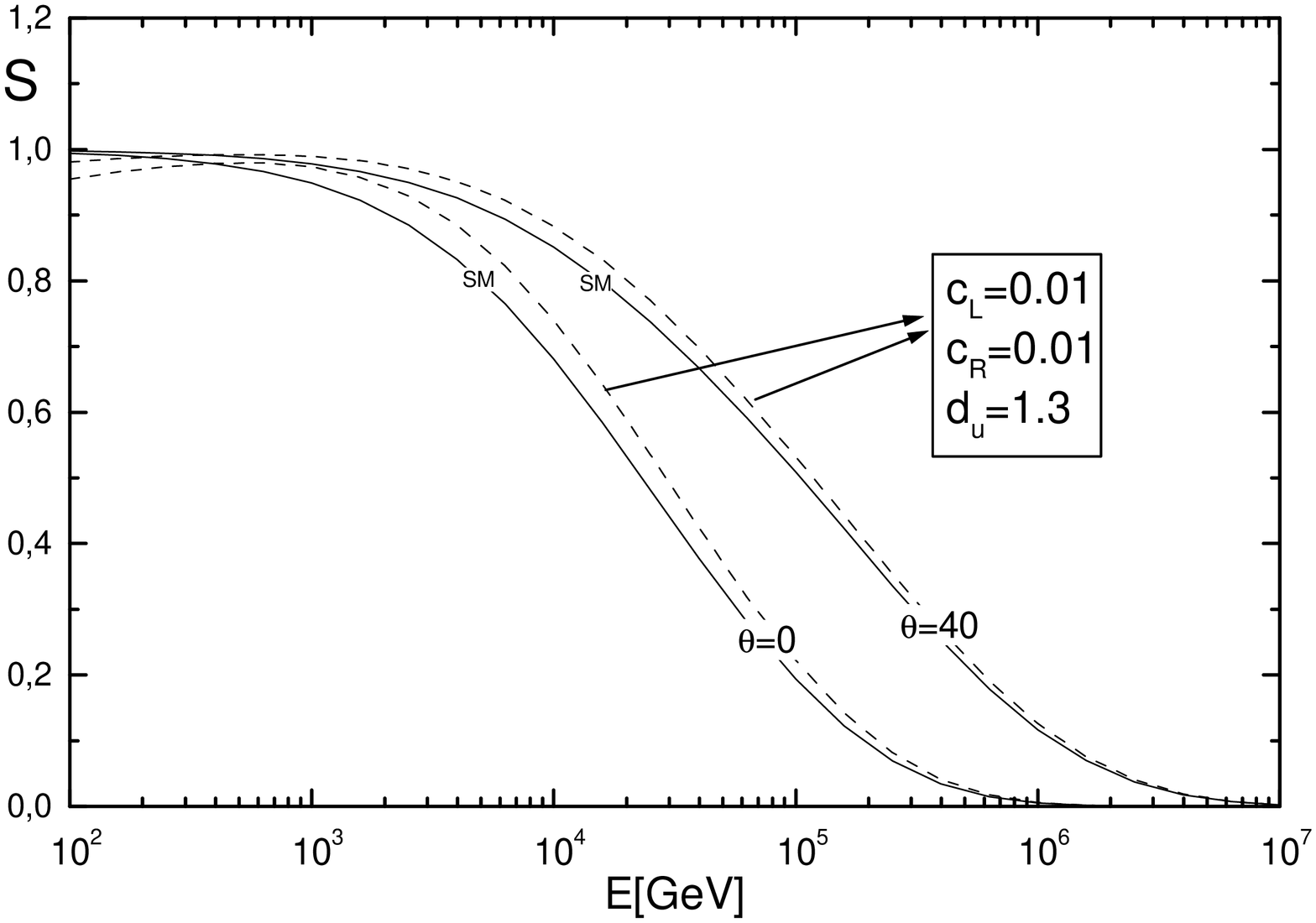}

\includegraphics[angle=270,width=8cm,bb= 150 180 650 680]{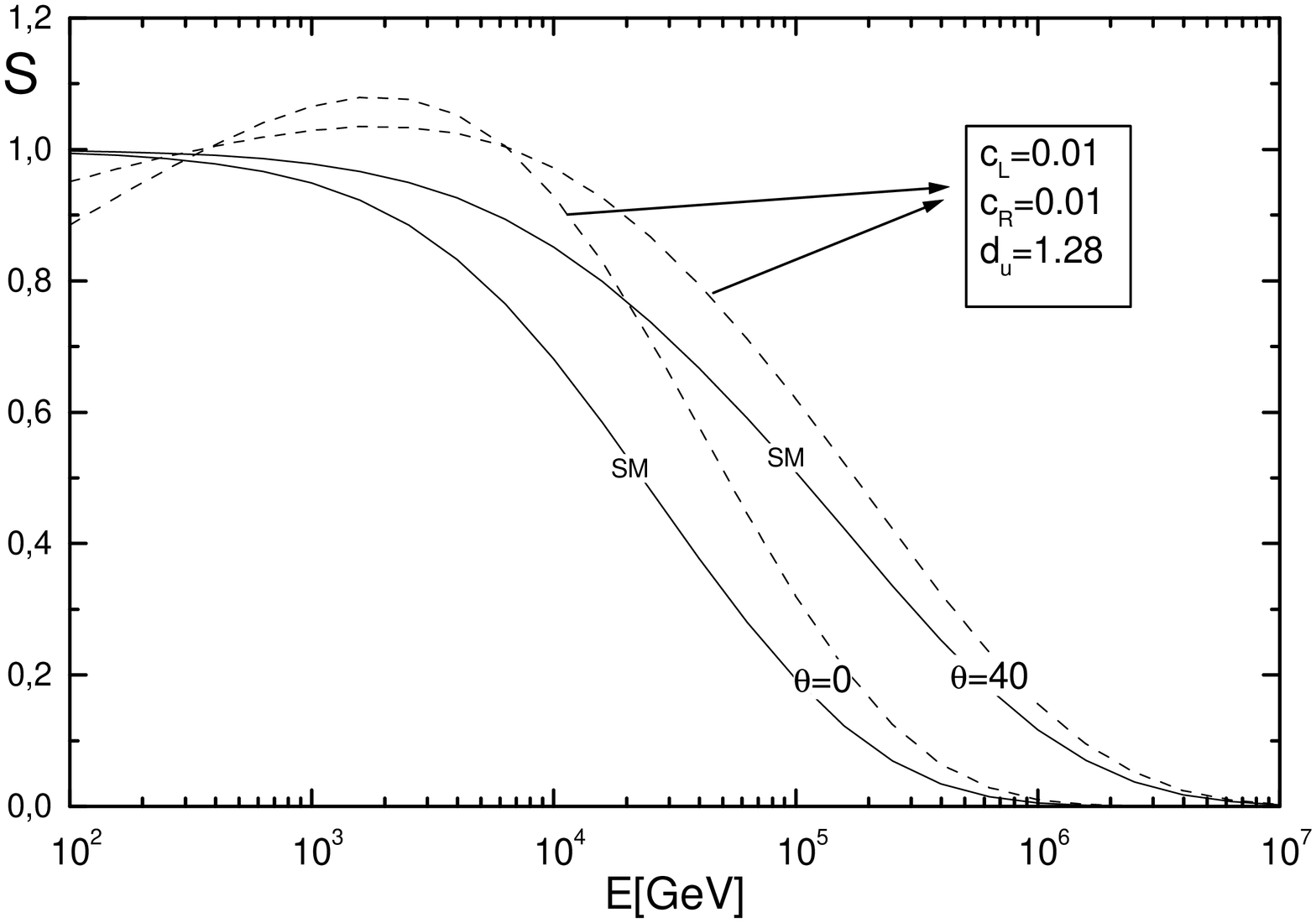}
\caption{\label{fig:SS}  $S$ for the SM and for different values of
$d_u$ at two different angles with respect to nadir. }
\end{figure*}

For  $d_u \geq 1.3$ the absorption and the regeneration
practically compensate each other and, then, the values of $S$
are very near to the SM values (Fig. \ref{fig:psi} Upper). For
$d_u<1.3$ the absorption and regeneration effects do not cancel
that efficiently and then we have values of $S$ that clearly differ
from the SM value. In Fig. \ref{fig:SS} we show $S$ for different
values of the unparticle parameters and for two angles between
the directions of the neutrino beam and the nadir.

\section{The Observable {\bf $\alpha(E)$}}

The angle $\alpha(E)$ and the related ratio $\eta(E)$ introduced in
Ref.\cite{alfa} are the observable that we shall use in this
paper in order  to study the impact of the unparticle physics
on neutrino detection in a neutrino telescope such as IceCube. By
definition $\alpha(E)$ is the angle that divides the Earth into two
homo-event sectors. When neutrinos traverse the planet in their
journey to the detector, they find different matter densities, and
then, different number of nucleons to interact with. In this
conditions, the number of neutrinos that finally arrive to the
detector depends on the arrival directions, indicated by the angle
$\theta$ with respect to the nadir direction. If we consider only
upward-going neutrinos of a given energy $E$, that is, the ones with
arrival directions $\theta$ such that $0<\theta<\pi/2$, there will
always exist an angle
 $\alpha(E)$ such that the number of events for
$0<\theta<\alpha(E)$ equals that for $\alpha(E)<\theta<\pi/2$.

Clearly, the value of $\alpha(E)$ is energy dependent. For low
energies, the cross section decreases and the Earth becomes
transparent to neutrinos. In this case $\alpha(E)\rightarrow \pi/3$
for a diffuse isotropic flux since this angle divides the hemisphere
into two sectors with the same solid angle. Obviously for extremely
high energies, where most neutrinos are absorbed,
$\alpha(E)\rightarrow \pi/2$, and for intermediate energies
$\alpha(E)$ varies accordingly between these limiting behaviors.

In order to define $\alpha(E)$ we consider the expected number of
events (muon tracks though charged currents $\nu_{\mu}N$
interactions) at IceCube in the energy interval $\Delta E$ and in
the angular interval $\Delta \theta$ that can be estimated as
\begin{equation}\label{numberevent}
{\mathcal N}=n_{\rm T} T \int_{\Delta\theta}\int_{\Delta E} d\Omega
dE_{\nu} \sigma^{CC}(E) \Phi(E,\theta),
\end{equation}
where $n_{\rm T}$ is the number of target nucleons in the effective
detection volume, $T$ is the running time, and $\sigma^{CC}(E)$ is
the charged neutrino-nucleon cross section. We take  the
detection volume for the events equal to the instrumented volume for
IceCube, which is roughly 1 km$^3$ and corresponds to $n_{\rm
T}\simeq 6 \times 10^{38}$. The function $\Phi(E,\theta)$  in
Eq.(\ref{numberevent}) is the survival flux which is the solution
Eq.(\ref{transporte}) of the complete transport equation
\cite{nicolaidis}.

\begin{figure}[!t]
\includegraphics[angle=270,width=3in,bb= 180 180 550 680]{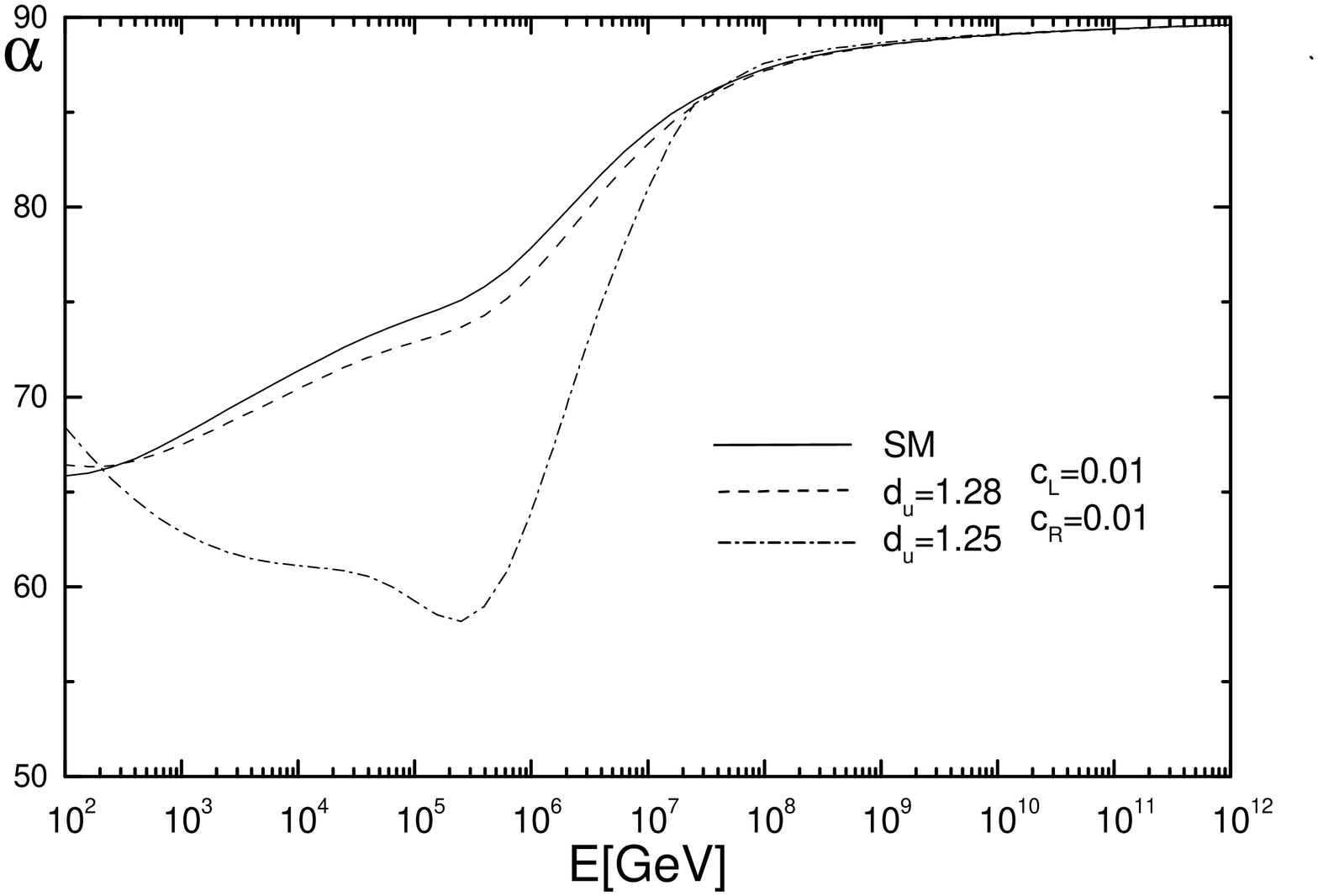}
\caption{\label{fig:alfunp}  The predictions for $\alpha(E)$
obtained for different values on $d_u$.}
\end{figure}

The definition of $\alpha(E)$ is essentially the equality between
two number of events, thus, to a good approximation, for each energy
bin all the previous factors cancel except the integrated fluxes at
each side. In this way, $\alpha(E)$ can be defined by the equation
\begin{equation}\label{alfadef}
\int_0^{\alpha(E)}d\theta \sin\theta e^{-\sigma_{eff}(E)
\tau(\theta)}=\int_{\alpha(E)}^{\pi/2} d\theta \sin\theta
e^{-\sigma_{eff}(E) \tau(\theta)},
\end{equation}
which is numerically solved to give the results shown in the
Fig.~\ref{fig:alfunp}. There we show the SM prediction for
$\alpha(E)$ and the unparticles contribution for different values of
the dimension $d_u$.

The main characteristics of $\alpha(E)$ have been reported recently
in Ref.\cite{alfa}. It is worth to notice that $\alpha(E)$ is weakly
dependent on the initial flux but, at the same time it is strongly
dependent on the neutrino nucleon cross-section. Hence, the use of
the observable $\alpha(E)$ reduces the effects of the experimental
systematics and initial flux dependence. Since the functional form
of $\alpha(E)$ sharply depends on the interaction cross section
neutrino-nucleon, if physics beyond the SM operates at
these high energies, it will become manifest directly onto the
function $\alpha(E)$.

In order to evaluate the impact of the observable $\alpha(E)$ to
bound new physics effects, we have estimated the corresponding
uncertainties on the SM prediction for $\alpha(E)$. Considering the
number of events as distributed according to a Poisson distribution
the uncertainty can be propagated onto the angle $\alpha_{\rm
SM}(E)$. The number of events $N$ as a function of $\alpha_{\rm SM}$
is
\begin{equation}
N=2 \pi n_{\rm T} T \Delta E \sigma^{CC}(E) \Phi_0(E)
\int_0^{\alpha_{\rm SM}}d\theta \sin\theta
e^{-\sigma_{eff}(E)\tau(\theta)},
\end{equation}
where we have considered the effective volume for contained events
so that an accurate and simultaneous determination of the muon
energy and shower energy is possible. For IceCube, it corresponds to
the instrumented volume, roughly 1 km$^3$, implying a number of
target nucleons $n_{\rm T}\simeq 6 \times 10^{38}$. We have
considered an integration time $T=15$ yr which is the expected
lifetime of the experiment.
To propagate the error on $N$ to obtain the one on $\alpha$, we note
that
\begin{equation}
\delta N=\dfrac{dN}{d\alpha} \delta\alpha,
\end{equation}
and dividing by $N$ we obtain 
\begin{equation}
\delta\alpha=\left[\int^{\alpha_{\rm SM}(E)}_0 d\theta \left(
\dfrac{\sin\theta}{\sin\alpha_{\rm SM}(E)} \right)
e^{\sigma_T(E)[\tau(\alpha_{\rm SM}(E))-\tau(\theta)]}\right]
\left( \dfrac{\delta N}{N} \right),
\end{equation}
where for Poisson distributed events we have
\begin{equation}
\delta N=\sqrt{N}.
\end{equation}

\begin{figure*}
\includegraphics[angle=270,width=2.7in,bb= 180 180 580 680]{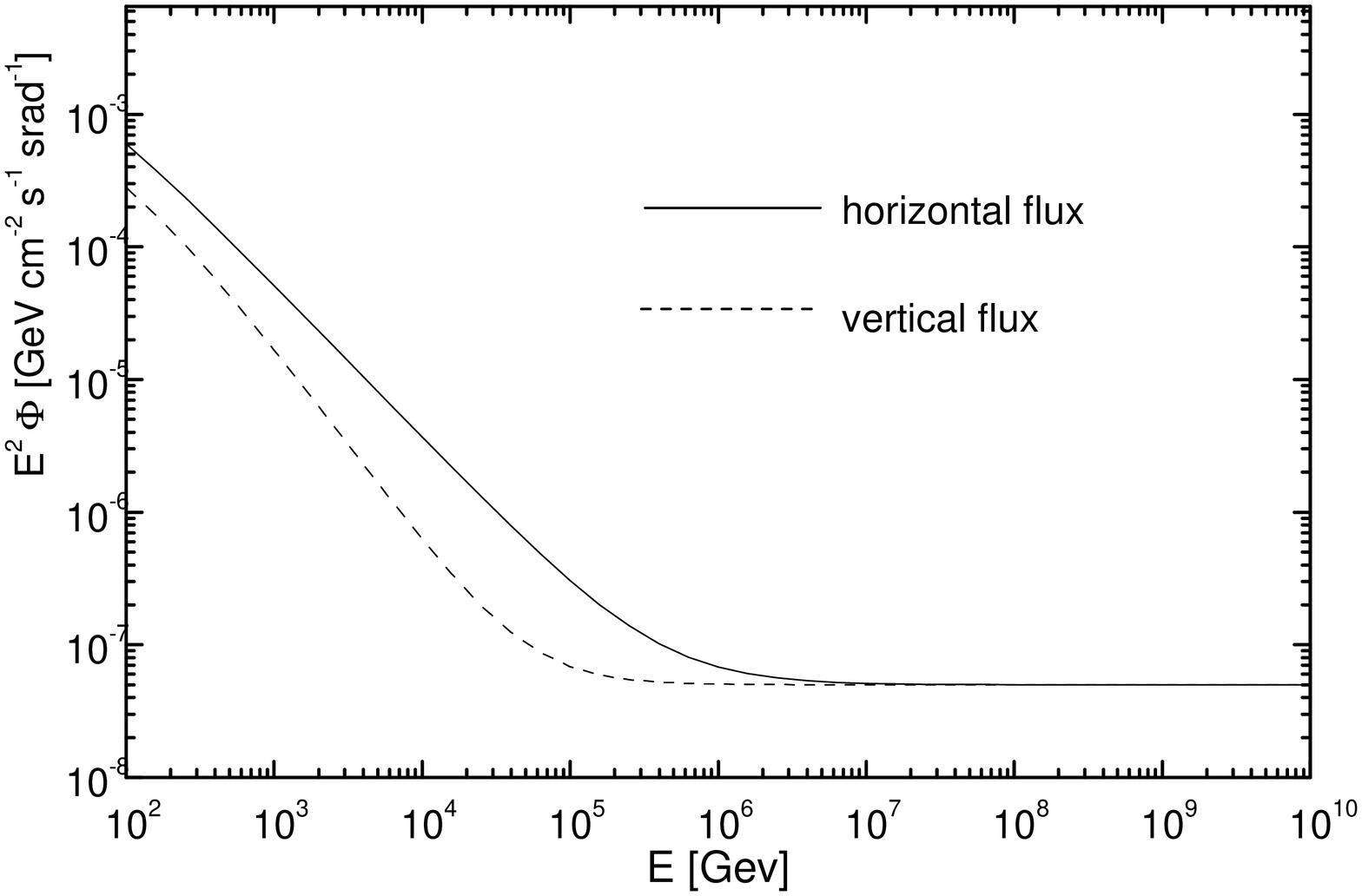}
\caption{\label{fig:flux0}  The utilized flux obtained by adding the
atmospheric and the isotropical cosmic flux. }
\end{figure*}

In order to evaluate the {\it errors} on $\alpha(E)$, it is
necessary to consider a level of initial flux $\Phi_0(E)$. Here we
have added together the cosmological diffuse isotropic flux and the
atmospheric one(see Fig.~\ref{fig:flux0}).  For the atmospheric
flux, we have considered the one given in Ref.\cite{atmos}. As for
the cosmological diffuse flux, the usual benchmark is the so-called
Waxman-Bahcall (WB) flux for each flavor, $E_{\nu_{\mu}}^2 \phi_{\rm
WB}^{\nu_{\mu}}\simeq 2.4\times 10^{-8} {\rm GeV} \ {\rm cm}^{-2}
{\rm s}^{-1} {\rm sr} ^{-1}$, which is derived assuming that
neutrinos come from transparent cosmic ray sources
\cite{waxman-bahcall}, and that there is an adequate transfer of
energy to pions following $pp$ collisions. However, one should keep
in mind that if there are in fact hidden sources which are opaque to
ultra-high energy cosmic rays, then the expected neutrino flux will
be higher.

On the other hand, we have the experimental bounds set by AMANDA. A
summary of these bounds can be found in
Refs.\cite{desiati,amanda-bound} and as a representative value we
take $E_{\nu_{\mu}}^2 \phi_{\rm AM}^{\nu_{\mu}}\simeq  8 \times
10^{-8} {\rm GeV} \ {\rm cm}^{-2} {\rm s}^{-1} {\rm sr} ^{-1}$. With
the intention to estimate the number of events, we have considered
an intermediate flux (INT) level slightly below the present
experimental bound by AMANDA,
\begin{equation}
E_{\nu_{\mu}}^2 \phi_{\rm INT}^{\nu_{\mu}}\simeq 5 \times 10^{-8}
{\rm GeV} \ {\rm cm}^{-2} {\rm s}^{-1} {\rm sr} ^{-1}.
\end{equation}

\begin{figure*}
\begin{center}
\includegraphics[angle=270,width=3in,bb= 100 180 680 680]{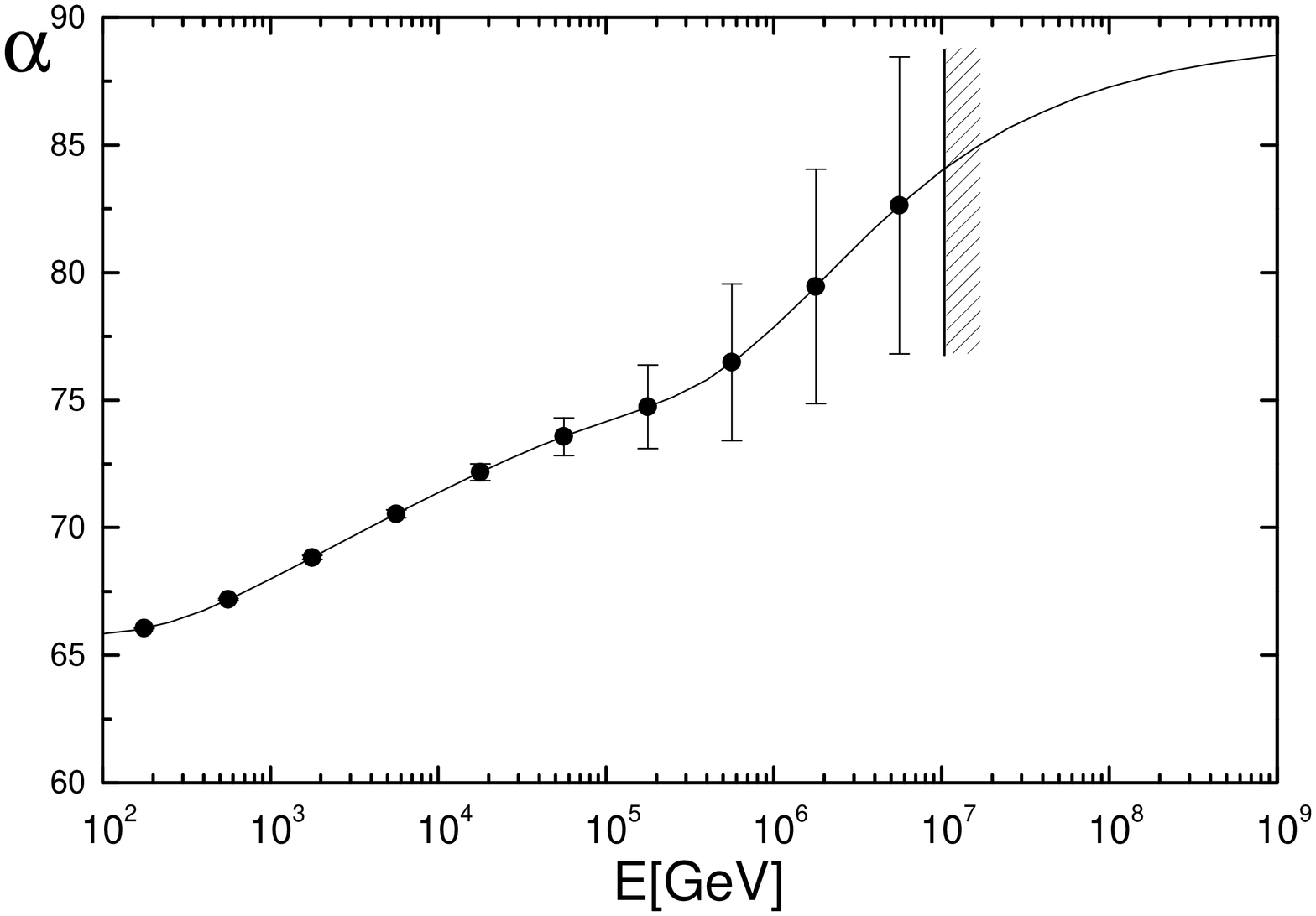} \caption{\label{fig:alferr}  The prediction for
$\alpha_{SM}(E)$ and the statistical errors.}
\end{center}
\end{figure*}

As it was discussed in Ref.\cite{alfa} the interval for maximum
sensitivity for $\alpha$ is $10^5 \rm{GeV} <E<10^7 \rm{GeV}$.
However, as for lower energies the atmospheric flux grows and then
the errors fall, we have considered as an energy window for the fits
the interval: $10^3 \rm{GeV} <E<10^7 \rm{GeV}$. In
Fig.~\ref{fig:alferr} we show our results for the observable
$\alpha$ and the corresponding errors within the mentioned energy
window. In Fig.~\ref{fig:flux0} we show the used flux.

\begin{figure*}
\centering
\includegraphics[scale=1.,angle=270,width=3.in,bb= 180 180 580 680]{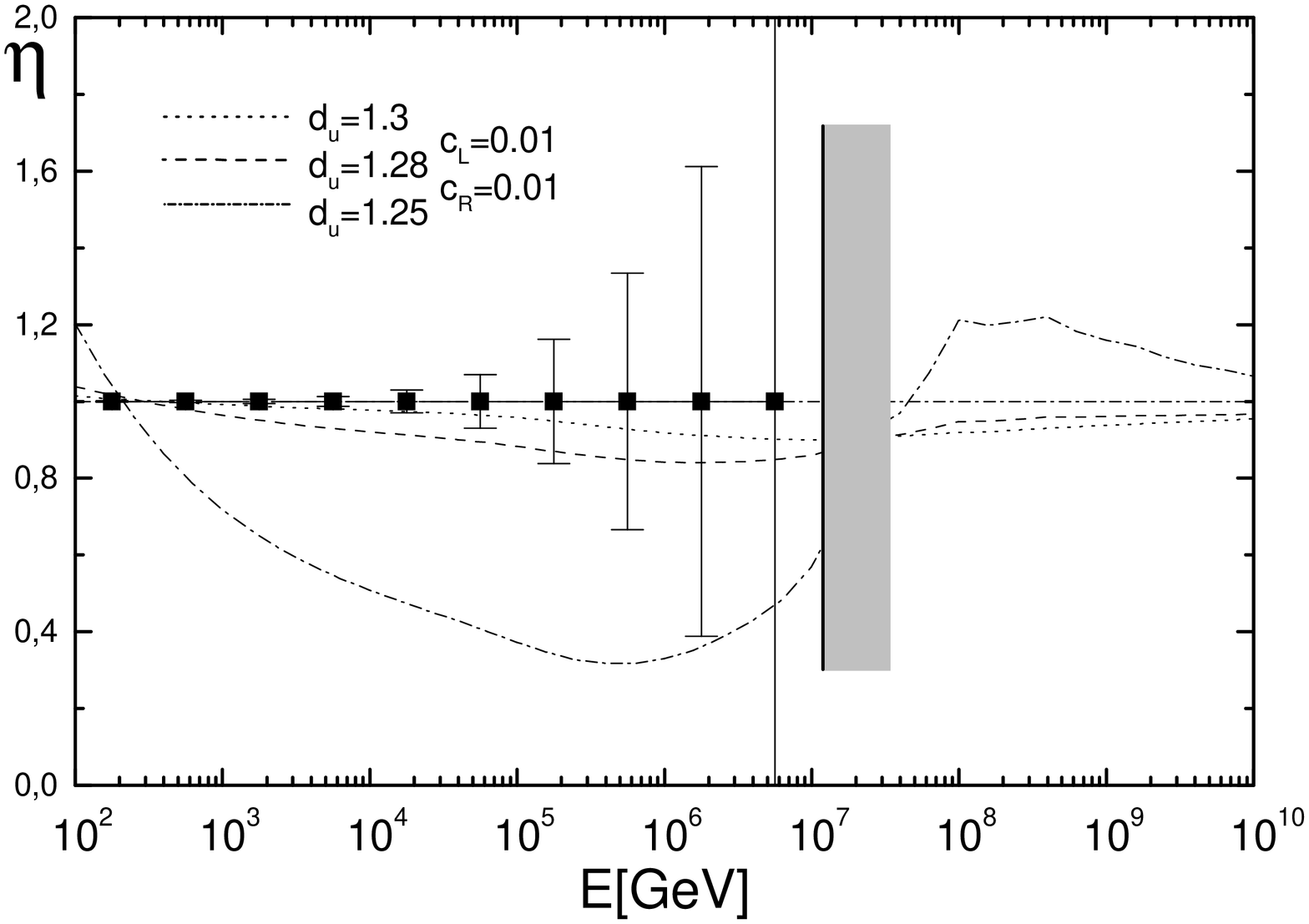}
\caption{\label{fig:capa} $\eta(E)$ for different values of $d_u$.
We include the statistical errors obtained of a number of events
distributed as a Poisson distribution.}
\end{figure*}

In the same context, we can define another observable related to
$\alpha(E)$. We consider the hemisphere $0<\theta<\pi/2$ divided
into two regions by the angle $\alpha_{SM}(E)$, ${\cal R}_1$ for
$0<\theta<\alpha_{SM}(E)$ and ${\cal R}_2$ for
$\alpha_{SM}(E)<\theta<\pi/2$. We then calculate the ratio $\eta(E)$
between the number of events for each region,
\begin{equation}\label{eta}
\eta(E)=\frac{N_1}{N_2},
\end{equation}
where $N_1$ is the number of events in the region ${\cal R}_1$ and
$N_2$ the number of events in the region ${\cal R}_2$. By using
$\eta(E)$ the effects of experimental systematic and initial flux
dependence are reduced. If there is only SM physics,
then we have that the ratio $\eta_{SM}(E)=1$. In order to estimate
the capability of $\eta(E)$ to bound unparticle effects, we have
considered the values of $\eta(E)$ along with their error bars in
Fig.~\ref{fig:capa} as if they had been obtained from experimental
measurements for $\eta(E)$. We proceed, then, to perform a
$\chi^2$-analysis taking as free parameters the dimension $d_u$ and
the constant $c=c_L=c_R$ and considering as {\it experimental point}
the SM values for $\eta(E)$ for the same energy bin used in
Fig.~\ref{fig:alferr}. We define the $\chi^2$ function in the usual
way,
\begin{equation}
\chi^2=\sum_{i=1,8} \dfrac{(\eta_{\rm
SM}(E_i)-\eta(E_i,d_u,c))^2}{(\delta\eta(E_i))^2}.
\end{equation}
where according to the definition of $\eta(E)$ (Eq.(\ref{eta})) the
statistical errors are given by $\delta\chi(E_i)=2/\sqrt{N_i}$ for
events distributed according to a Poisson distribution. The function
$\chi^2$ is minimized to obtain the allowed region in the ($d_u$,
$c$) plane, which corresponds to the region below the curve {\bf A}
shown in Fig.~\ref{fig:capareg}. In the same figure we also include
other bounds obtained from different processes. The curve {\bf B}
corresponds to bounds obtained from atomic parity
violation\cite{vap}, the curve {\bf C} corresponds to bounds from
the muon anomaly \cite{am} and curve {\bf E} comes from low energy
$\nu_e - e$  scattering\cite{nes}. This last bound is
significatively better than all the other bounds. We would like to
stress that the curve {\bf E} corresponds to $\nu_e - e$ scattering
but we are considering muon neutrinos, i.e., a different neutrino
flavor. Thus, if we consider the unparticle interactions as
non-universal then this constraint would not apply and the neutrino
telescope bounds may be competitive. However, there are experimental
data for the ratio between $\nu_{\mu}-N$ Neutral Current to Charged
Current interaction $R$ \cite{Rexp}:

\begin{equation}
R(E_{\nu_{\mu}})=\frac{\sigma^{NC}_{\nu N\rightarrow \nu
X}(E)}{\sigma^{CC}_{\nu N\rightarrow \mu X}(E)}
\end{equation}

In order to compare with the experimental value

\begin{equation}\label{rexperi}
R=0.320\pm 0.010 ,
\end{equation}

which is an average measurement for energies in the range  20 GeV -
200 GeV, we calculated the corresponding average value for $R$:

\begin{equation}
\langle R \rangle=\frac{1}{180 GeV}\int_{20GeV}^{200 GeV} R(E) dE
\end{equation}

and compared it with the previous experimental value
(Eq.(\ref{rexperi})) for different values of the unparticles
parameters ($d_u,c$). The allowed region corresponds to the points
below the curve {\bf D} of Fig.~\ref{fig:capareg}.  This curve shows
that the bound coming from R is stronger than the one IceCube could
possible set in the future.

\section{Conclusions}
In the present work we studied the effects of unparticles
contributions to the neutrino-nucleon cross section on the survival
neutrino flux in a neutrino telescope like IceCube. To do it, we
 considered an effective interaction between standard particles
and unparticles. We have  found a considerable disagreement with the
SM prediction for the neutrino observables defined above,
particulary for low values of $d_u$ and low neutrino energy. For
moderate values of $d_u$ this disagreement tends to disappear due to
a strong cancellation between absorption and regeneration.

\begin{figure*}
\begin{center}
\includegraphics[scale=1.,angle=270,width=3.in,bb= 100 180 680 680
]{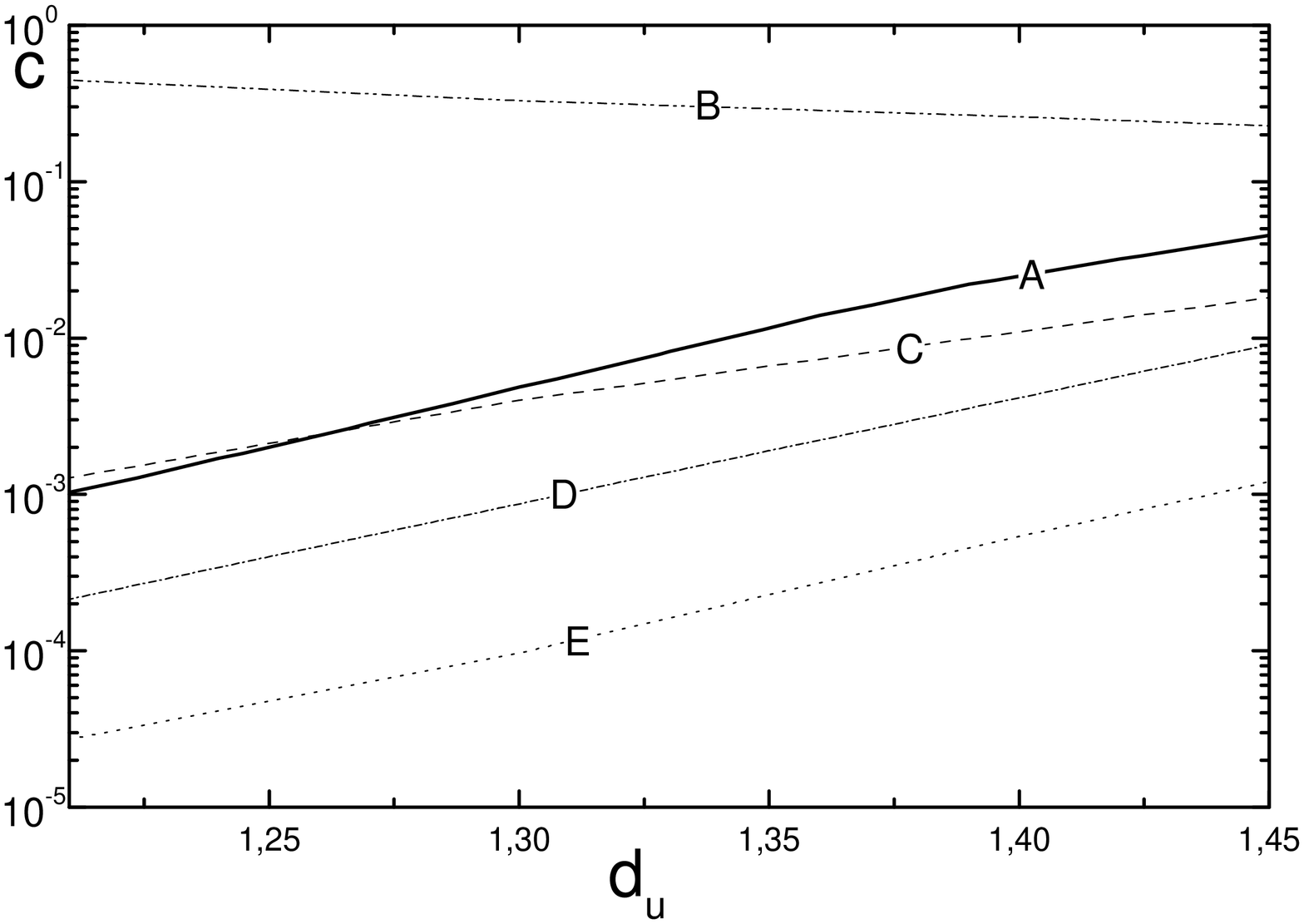} \caption{\label{fig:capareg} The allowed
regions in the $(d_u,c)$ plane are below the corresponding curves.
Curve A: this paper, Curve B: bounds obtained form Atomic Parity
Viloation \cite{vap}, Curve C: it obtained from $(g-2)_{\mu}$
\cite{am}, Curve E correspond to bounds obtained from  $\nu_e - e$
scattering \cite{nes}, and curve {\bf D} correspond to ones obtained
from the ratio $\langle R \rangle$ \cite{Rexp}.}
\end{center}
\end{figure*}

We have  also studied the possibility to bound effects of
unparticles contributions to the interactions between muon neutrinos
and the nucleons of the Earth using the observable $\eta(E)$.  In
this context, we fitted the theoretical expression for $\eta(E)$ as
a function of the $d_u$ and $c$ taking as experimental data the SM
values obtained for $\eta$ ($\eta_{SM}(E)=1$) along with the errors
derived for a number of events distributed according to a Poisson
distribution. The results are shown in Fig.~\ref{fig:capareg} as a
allowed region plot. Finally, in the same figure we have compared
the limits that have been obtained  for different
  authors for low energy processes. The bounds coming from the ratio R of neutral to charged currents
(curve D) is more restrictive than the bound IceCube could set
(curve A) and then new effects that  IceCube may detect could not be
attributable to unparticle physics.

\begin{acknowledgments}
We thank CONICET (Argentina) and Universidad Nacional de Mar del
Plata (Argentina), the  Fundaci\'on del Banco de la Republica
(Colombia), PDT 54/094 (Uruguay) and CLAF for the financial
supports.
\end{acknowledgments}


\end{document}